\documentclass[12pt,a4paper]{article}
\usepackage{graphicx}
\usepackage{epsfig}

\begin{document}

\begin{center}
 {\bf \Large $J$-matrix method and Bargmann  potentials}\\[3pt]
{\large  S.~A.~Zaitsev, E.~I.~Kramar}\\[2pt]
{\sl Department of Physics, Khabarovsk State Technical University,\\
Tikhookeanskaya 136, Khabarovsk 680035, Russia\\
 E-mail: zaytsev@mail.khb.ru}\\[4pt]
\end{center}
\begin{abstract}
By applying the $J$-matrix method \cite{Jmtx} to neutral particles scattering
we have discovered that there is a one-to-one correspondence between the
nonlocal separable potential with the Laguerre form factors and a Bargmann
potential. Thus this discrete approach to direct and inverse scattering
problem can be considered as a tool of the $S$-matrix rational parametrization.
As an application, the Bargmann potentials, phase-equivalent to the $np$
$^1S_0$ Yamaguchi potential \cite{Yamaguchi} and to the $np$ potential from
inverse scattering in the $J$-matrix approach \cite{ZInv} have been obtained.
\end{abstract}

\section{Introduction}
\par In the $J$-matrix method \cite{Jmtx} the initial short-ranged interaction
is approximated by the model nonlocal separable potentials with the harmonic
oscillator or Laguerre form factors. In the latter case the $S$-matrix is a
rational function of the wave number $k$. This allows to establish the
correspondence between the model potential and the phase-equivalent Bargmann
potential \cite{Newton,CS}. Bargmann's type potentials were used extensively
for nuclear systems description (see e. g. \cite{Zakhariev} and references
therein). Recent results see in Ref. \cite{SB}. As a rule the model potentials
of rather high rank $N$ ($N \geq 10$) are necessary for the $J$-matrix
description for the initial interaction. This makes solving the Marchenko
integral equation for the Bargmann potentials difficult. On the other hand, the
inverse scattering method within the $J$-matrix approach \cite{ZInv} enables
one to construct the model potentials of low rank. In view of noticed the
$S$-matrix property of the model potential, this inverse scattering method is
at once an efficient tool for the Bargmann potentials construction.
\par In Sec.~2 the $S$-matrix of the separable potential with the Laguerre form
factors was demonstrated to be a rational function of the wave number $k$. In
Sec.~3 for the Bargmann potential Jost function in terms of the results of
inversion procedure within the $J$-matrix approach \cite{ZInv} is given in an
explicit form. In Sec.~4 the Bargmann potentials have been obtained,
phase-equivalent to the $np$ $^1S_0$ potentials: (i) Yamaguchi of rank 1
potential \cite{Yamaguchi}; (ii) separable potential of rank 4 from inverse
scattering in the $J$-matrix method. In Sec. 5 we summarize our conclusions.

\section{The model separable potentials}
\par In the $J$-matrix approach \cite{Jmtx} the short-ranged potential in the
 partial wave $\ell$ is approximated by a separable expansion:
\begin{equation}
\widehat{V}^\ell=\frac{\hbar^2}{2\mu} \sum_{n, \, n' =0}^{N-1} \left|
\overline{\phi}_n^{\ell}\right> V_{n, \, n'} \left<
\overline{\phi}_{n'}^{\ell}\right|, \label{PPot}
\end{equation}
where the form factors
\begin{equation}
  \left| \overline{\phi}_n^{\ell} \right>= \frac{\displaystyle n!}
  {\displaystyle r \, (n+2\ell+1)!} (2 b r)^{\ell+1} \,
  e^{-b r} L_n^{2\ell+1}(2 b r) \label{orthbf}
\end{equation}
are bi-orthogonal to the Laguerre basis functions
\begin{equation}
  \left| \phi_n^{\ell} \right> = (2 b r)^{\ell+1} \
  e^{-b r} L_n^{2\ell+1}(2 b r), \label{basfun}
\end{equation}
i. e.
\begin{equation}
\int \limits _0 ^{\infty} \phi_n^{\ell}(r)\,
\overline{\phi}_{n'}^{\ell}(r)\,dr= \delta_{n \, n'}.
\end{equation}
Here, $b$ is a scaling parameter.
\par In the neutral particles scattering case the Fredholm determinants
${\cal D}^{(\pm)}(k)$ corresponding to the Lippman-Schwinger equation solutions
is given by (see e. g. \cite{Arnold}):
\begin{equation}
{\cal D}^{(\pm)}(k) \equiv \det({\bf I}- {\bf G}^{(0)(\pm)}{\bf V}). \label{D1}
\end{equation}
Here, $[{\bf V}]_{n, \, n'}=V_{n, \, n'}$ are the matrix elements of the
potential (\ref{PPot}); the matrix elements of the free Green's operator
$\widehat{G}^{(0)(\pm)}(k)=$ $\left[k^2 \pm \mbox{i}\varepsilon-\widehat{h}_0
\right]^{-1}$ ($\widehat{h}_0 = -\frac{\displaystyle d^2}{\displaystyle
d\,r^2}+ \frac{\displaystyle \ell(\ell+1)}{\displaystyle r^2}$) are defined by
the expressions \cite{Broad,YA}:
\begin{equation}
[{\bf G}^{(0)(\pm)}]_{n, \, n'}= \langle \overline{\phi}_n^{\ell} \left|
\widehat{G}^{(0)(\pm)}(k) \right| \overline{\phi}_{n'}^{\ell}\rangle=
-\frac{1}{k} \,S_{n_{<}}^{\ell}(k)C_{n_{>}}^{\ell(\pm)}(k) \label{Green}
\end{equation}
where
\begin{equation}
 \begin{array}{c}
C_n^{\ell(\pm)}(k) = - \frac{\displaystyle n! \,  e^{\mp \mbox{i}(n+1) \,
\theta } } { \displaystyle (n+\ell+1)! \left(2\, \sin \theta \right)^{\ell}}\;
{_2F_1}(-\ell, \,
n+1; \, n+\ell+2; \, e^{\mp 2\mbox{i}\theta}), \\[5mm]
S_n^{\ell}(k) = \frac{\displaystyle \ell! \left(2\, \sin \theta
\right)^{\ell+1} } {
\displaystyle \vphantom{{I^I}^I} 2 \, (n+2\ell+1)!}\; {\cal C}_n^{\ell+1}(\cos \theta), \\[5mm]
e^{\mbox{i} \theta}=(k+\mbox{i}b)/ (k-\mbox{i}b),
 \end{array}
\label{CS}
\end{equation}
${\cal C}_n^{\ell+1}(x)$ represent the Gegenbauer polynomials. Reference to
Eqs. (\ref{Green}), (\ref{CS}), shows that the Fredholm determinants ${\cal
D}^{(\pm)}(k)$ (\ref{D1}) are rational functions of the wave number $k$,
furthermore, as shown below, ${\cal D}^{(\pm)}(k)$ can be written in the form:
\begin{equation}
{\cal D}^{(\pm)}(k) = \frac{\prod \limits_{j=1}^{\mathcal{N}} (k \pm
\mbox{i}a_j) } {\vphantom{{I^I}^I} (k \pm \mbox{i} b)^{\mathcal{N}}}, \qquad
\mathcal{N}=2(N+\ell). \label{D12}
\end{equation}
Notice that since $S$-matrix of the potential (\ref{PPot}) is defined by
\cite{Arnold}
\begin{equation}
S(k) = \frac{\displaystyle {\cal D}^{(-)}(k)}{\displaystyle {\cal D}^{(+)}(k)},
\end{equation}
it is also rational function of $k$:
\begin{equation}
S(k) = \left(\frac{\displaystyle k+\mbox{i} b}{\displaystyle k-\mbox{i}
b}\right)^{\mathcal{N}} \prod \limits_{j=1}^{\mathcal{N}} \frac{
(k-\mbox{i}a_j) }{(k+\mbox{i}a_j) }. \label{Sm}
\end{equation}
Consequently, Eq. (\ref{Sm}) simultaneously represents the $S$-matrix of a
Bargmann potential, phase-equivalent to the potential (\ref{PPot}).
\par Let us next assume that the Bargmann potential Jost function $F(k)$
coincides with the Fredholm determinant $\mathcal{D}^{(+)}(k)$ of the
potential (\ref{PPot}):
\begin{equation}
F(k)=\frac{\prod \limits_{j=1}^{\mathcal{N}} (k +\mbox{i}a_j) }
{\vphantom{{I^I}^I} (k+\mbox{i} b)^{\mathcal{N}}}.  \label{Jf}
\end{equation}
The parameters $\left\{ a_j \right\}$ must obey the following restrictions:
either $a_j < 0$ corresponding to the introduction of new bound state, or $Re
\,a_j>0$. In the latter case the numerator in Eq. (\ref{Jf}) should contain
the factor $k+\mbox{i}a_j^*$. So that in the absence of bound states the
$S$-matrix has only pole in the upper half $k$ plane, namely, in the point
$\mbox{i}b$. This implies that, for instance, in the $s$-wave case the kernel
$Q(r, \, r')=Q(t)$ ($t=r+r'$) of the Marchenko integral equation
\cite{Zakhariev} is of the form
\begin{equation}
Q(t) = -\mbox{i}\, \mathop{Res} \limits_{k= \mbox{i}b} \left\{ S(k)\,
e^{\mbox{i} k t}\right\},
\end{equation}
i. e. the separable kernel $Q(t)$ can be represented in the form of the sum of
$\mathcal{N}$ components:
\begin{equation}
 \begin{array}{c}
Q(r, \, r')= \sum \limits_{n=0}^{\mathcal{N}-1} Q^{(1)}_n(r)\,
Q^{(2)}_n(r'),\\[3mm]
 Q^{(1)}_n(r)= r^n \, e^{-b r}, \qquad Q^{(2)}_n(r)=e^{-b r} \sum \limits
 _{m=n} ^{\mathcal{N}-1} A_m {m \choose n} \, r^{m-n},\\[3mm]
  A_m = - \frac{\displaystyle (\mbox{i})^{(m+1)}}{\displaystyle
 {\phantom{C^C}  m!\,(\mathcal{N}-m-1)!}}\, \left\{
(k-\mbox{i}b)^{\mathcal{N}} S(k)\right\}^{(\mathcal{N}-m-1)}
 \Biggl|_{k=\mbox{i}b} \Biggr. .\\
 \end{array}
\end{equation}

\section{Fredholm determinant in the $J$-matrix method}
\par The general solution for multichannel Fredholm determinant for the model
potential has been obtained in \cite{YA}. Here, we only express the
single-channel Fredholm determinant in terms of the scattering inversion
\cite{ZInv} results: the eigenvalues $\left\{ \lambda_j \right\}$ and the
elements $\left\{ Z_{N, \, j} \right\}$ of the eigenvectors orthogonal matrix
${\bf Z}$ of the $N \times N$ Hamiltonian $\widehat{h}=$ $
\widehat{h}_0+\frac{2\mu}{\hbar^2}\widehat{V}^\ell$ matrix calculated in the
orthonormalized basis $\left\{ \left| \psi_n^\ell \right> \right.$, $\left.
n=\overline{0, \; N-1} \right\}$:
\begin{equation}
\left| \psi_n^\ell \right>=d_n \, (2 b r)^{\ell+1}\, e^{-b r} \,
L_n^{2\ell+2}(2 b r), \qquad d_n = \sqrt{\frac{\displaystyle 2 b n!
}{\displaystyle (n+2\ell+2)!}}. \label{od}
\end{equation}
For this purpose the solution for the $S$-matrix in the framework of the
$J$-matrix method \cite{YA}:
\begin{equation}
S(k) = \frac{\displaystyle C_{N-1}^{\ell(-)}(k)-{\cal P}_N(k^2) J_{N-1, \,
N}(k) \,C_{N}^{\ell(-)}(k) } {\displaystyle C_{N-1}^{\ell(+)}(k)-{\cal
P}_N(k^2) J_{N-1, \, N}(k) \,C_{N}^{\ell(+)}(k)} \label{S2}
\end{equation}
is used. Here, $J_{n, \,n'}(k)$ are the elements of the tridiagonal $J$-matrix:
\cite{Jmtx,YA}
\begin{equation}
 \begin{array}{c}
J_{n, \, n}(k) = \frac{\displaystyle 1}{\displaystyle b}(n+\ell+1)
\frac{\displaystyle (n+2\ell+1)!}{\displaystyle n!}(b^2-k^2),\\[3mm]
J_{n, \, n+1}(k) = J_{n+1, \, n}(k)=\frac{\displaystyle
(n+2\ell+2)!}{\displaystyle 2\, b \, n!}
(b^2+k^2);\\
 \end{array}
 \label{Jm}
\end{equation}
The ${\cal P}$-matrix ${\cal P}_{N}(k^2)$ is defined by \cite{BR,ZInv}
\begin{equation}
{\cal P}_{N}(k^2)= d_{N-1}^2 \, \sum_{j=1}^{N} \frac{\displaystyle Z_{N, \,
j}^2}{\displaystyle k^2-\lambda_j}. \label{Pm}
\end{equation}
\par Using the expression of so-called Casoratian determinant \cite{BR,YA}:
\begin{equation}
C_{n+1}^{\ell(+)}(k)\,S_{n}^{\ell}(k)-C_{n}^{\ell(+)}(k)\,S_{n+1}^{\ell}(k)=
\frac{\displaystyle k}{\displaystyle J_{n, \, n+1}(k)},
\end{equation}
in view of three-term recursion relations \cite{Jmtx}:
\begin{equation}
 \begin{array}{c}
 J_{n, \, n-1}(k)\,S_{n-1}^{\ell}(k)+J_{n, \, n}(k)\,S_{n}^{\ell}(k)
 +J_{n, \, n+1}(k)\,S_{n+1}^{\ell}(k)=0, \; n>0,\\[3mm]
 J_{0, \, 0}(k)\,S_{0}^{\ell}(k)
 +J_{0, \, 1}(k)\,S_{1}^{\ell}(k)=0,\\[3mm]
 J_{n, \, n-1}(k)\,C_{n-1}^{\ell(+)}(k)+J_{n, \, n}(k)\,C_{n}^{\ell(+)}(k)
 +J_{n, \, n+1}(k)\,C_{n+1}^{\ell(+)}(k)=0, \; n>0,\\[3mm]
 J_{0, \, 0}(k)\,C_{0}^{\ell(+)}(k)
 +J_{0, \, 1}(k)\,C_{1}^{\ell(+)}(k)=
 \frac{\displaystyle k}{\displaystyle S_{0}^\ell(k)}, \\[3mm]
 \end{array}
\end{equation}
the relation between the Fredholm determinant ${\cal D}^{(+)}(k)$ and the
denominator of the expression (\ref{S2}) of the $S$-matrix can be developed:
\begin{equation}
{\cal D}^{(+)}(k)=-\frac{\displaystyle S_0^\ell(k)}{\displaystyle k} \, J_{N-1,
\, N}(k) \left\{ C_{N-1}^{\ell(+)}(k)-{\cal P}_N(k^2) J_{N-1, \, N}(k)
\,C_{N}^{\ell(+)}(k) \right\} \prod_{j=1}^{N}\frac{\displaystyle
(k^2-\lambda_j)}{\displaystyle d_{j-1}^2 \, J_{j-1, \, j}(k)} , \label{D2}
\end{equation}
Using equations (\ref{CS}), (\ref{od}) and (\ref{Jm}) for the solutions
$S_n^\ell(k), \, C_{n}^{\ell(+)}(k)$, the normalization factors $d_n$ and the
elements $J_{n, \, n+1}(k)$, respectively, the expression (\ref{D2}) can be
rewritten as
\begin{equation}
 \begin{array}{l}
{\cal D}^{(+)}(k)= \frac{\displaystyle 1}{\displaystyle
\vphantom{{I^I}^I}(k+\mbox{i} b)^{2N} } \, \frac{\displaystyle \ell! \,
(N+2\ell+1)!}{\displaystyle \vphantom{{I^I}^I} (2\ell+1)!
\, (N+\ell+1)!} \\[5mm]
\phantom{CC} \left\{ (N+\ell+1) \, {_2 F _1}\left(-\ell,\, N \; ; N+\ell+1 \; ;
\left(\frac{\displaystyle k-\mbox{i}b}{\displaystyle k+\mbox{i}b} \right)^2
\right)
\prod \limits_{j=1}^N (k^2-\lambda_j)-  \right. \\[5mm]
\phantom{CCC} \left. -N \,(k-\mbox{i} b)^2 \,{_2 F _1}\left(-\ell,\, N+1 \; ;
N+\ell+2 \; ; \left(\frac{\displaystyle k-\mbox{i}b}{\displaystyle k+\mbox{i}b}
\right)^2 \right) \sum \limits_{j=1}^{N} \left[ Z_{N, \, j}^2 \prod
\limits_{i=1 \atop {i \neq j}}^{N}
(k^2-\lambda_i) \right]\right\}.\\
 \end{array}
\label{F2}
\end{equation}
Since hypergeometric functions ${_2 F _1}$ in the right-hand side of equation
(\ref{F2}) are polynomials of degree $\ell$ in $\left(\frac{\displaystyle
k-\mbox{i}b}{\displaystyle k+\mbox{i}b} \right)^2$, ${\cal D}^{(+)}(k)$
represents a rational polynomials in $k$ of degree $\mathcal{N}$:
\begin{equation}
{\cal D}^{(+)}(k)=
\frac{\mathcal{R}_{\mathcal{N}}(k)}{\vphantom{{I^I}^I}(k+\mbox{i}
b)^{\mathcal{N}}}.
\end{equation}
Notice that the leading coefficient of the polynomial
$\mathcal{R}_{\mathcal{N}}(k)$ is, obviously, equal to
\begin{equation}
 \begin{array}{l}
\frac{\displaystyle \ell! \, (N+2\ell+1)!}{\displaystyle \vphantom{{I^I}^I}
(2\ell+1)! \, (N+\ell+1)!}
\Bigl\{ (N+\ell+1) \, {_2 F _1}\left(-\ell,\, N \; ; N+\ell+1 \; ; 1 \right) \Bigr.\\
\phantom{CCCCCCCCCCCCCCCCCCCC} \Bigl. -N \,{_2 F _1}\left(-\ell,\, N+1 \; ;
N+\ell+2 \; ; 1 \right) \sum \limits_{j=1}^{N} Z_{N, \, j}^2 \Bigr\}=1,
\end{array}
\end{equation}
since ${_2 F _1}\left(a,\, b \; ; c \; ; 1 \right)=$ $\frac{\displaystyle
\Gamma(c)\Gamma(c-a-b) }{\displaystyle  \vphantom{{I^I}^I}
\Gamma(c-a)\Gamma(c-b)}$, and from the orthogonality of ${\bf Z}$ follows $\sum
\limits_{j=1}^{N} Z_{N, \, j}^2=1$. Thus equation (\ref{D12}) is proved.

\section{Examples}
\par As a first example, the Bargmann potential has been obtained,
phase-equivalent to the $np$ $^1S_0$ Yamaguchi potential \cite{Yamaguchi}:
\begin{equation}
V(k, k')= \Lambda_0 \frac{\displaystyle 1}{\displaystyle \vphantom{{I^I}^I}
(k^2+b^2)} \frac{\displaystyle 1}{\displaystyle
\vphantom{{I^I}^I}({k'}^2+b^2)}, \label{Ymg}
\end{equation}
where $\Lambda_0=-76,4294$ $MeV \cdot fm^{-1}$, $b=1,158023$ $fm^{-1}$. The
potential (\ref{Ymg}) corresponds to the potential (\ref{PPot}) of the rank
$N=1$ where $V_{0,\,0}= \frac{\displaystyle 2 \mu}{\displaystyle
\hbar^2}\frac{\displaystyle \pi}{\displaystyle \vphantom{I^I}8b^2}\Lambda_0$.
The Jost function (\ref{Jf}) parameters $b$, $\left\{ a_j \right\}$ are listed
in Table. The corresponding Bargmann potential is displayed in Fig.~1.
\par In a second example the potential (\ref{PPot}) of rank $N=4$ has been
constructed by the inverse scattering method \cite{ZInv} within the $J$-matrix
approach. In doing this a set of Nijmegen \cite{Nijmegen} phase shifts in the
$0-350$ $MeV$ energy range is taken as input. The scaling parameter $b$ and
the set $\left\{ \lambda_i, \, Z_{4, \, i} \right.$, $\left. i=\overline{1, \;
4}\right\}$ are given in Table determine the $N \times N$ matrix of the
potential (\ref{PPot}). $\mathcal{N}=8$ parameters $\left\{ a_j \right\}$ of
the Bargmann potential Jost function (\ref{Jf}) are listed in Table. Structure
of deep and shallow parts of the Bargmann potential is shown in Figs.~2a-2d.

\section{Conclusion}
\par In summary, a neutral particle scattering from nonlocal separable
potential with the Laguerre form factors has been dealt with in this work. For
the potential (\ref{PPot}) of rank $N$ in the partial wave $\ell$ the Fredholm
determinant was demonstrated to represent a rational polynomial in $k$ of
degree $\mathcal{N} =2(N+ \ell)$. Thus a correspondence between the potential
(\ref{PPot}) and the Bargmann potential may be established by identifying the
Jost function of the latter with the Fredholm determinant of the model nonlocal
potential. Here, the phase-equivalence between the potential (\ref{PPot}) and
its Bargmann's analogue is exact. As one goes to the Coulomb reference
Hamiltonian $\widehat{h}_0$ or to the harmonic oscillator form factors case
the Eq. (\ref{D2}) remains valid. However, the solutions $S_n^\ell(k), \;
C_{n}^{\ell(+)}(k)$ are not longer rational functions of the wave number $k$
\cite{YA}. In this case Bargmann potential is only approximately
phase-equivalent to the initial separable potential, since here the Fredholm
determinant is approximated by rational expression (\ref{D12}). The parameters
$\left\{ a_j \right\}$ coincide with $\mathcal{N}$ (multiplied by imaginary
unity) roots of the expression enclosed in figured parentheses in the Eq.
(\ref{D2}).

\subsection*{Acknowledgment}
\par It is pleasure to thank Prof. A~.M.~Shirokov and Prof. A.~I.~Mazur for
helpful discussions. This work was supported in part by the State Program
``Universities of Russia'', project No 992306.

\newpage
Table. The parameters values of the $np$ $^1S_0$ Bargmann potentials,
phase-equivalent to: (i) the Yamaguchi potential (\ref{Ymg}) of rank 1
\cite{Yamaguchi}; (ii) the potential (\ref{PPot}) of rank 4 from inverse
scattering in the $J$-matrix approach. The parameters set $\left\{
\lambda_i,\right.$ $\left.Z_{4, \, i} \right.$, $\left. i=\overline{1, \;
4}\right\}$ of the separable potential is also presented here.

{\small
$$
\begin{array}{cc|ccccc}
\hline \hline
 & & & & & & \\
\multicolumn{2}{c}{ b=1.158023 \, fm^{-1}}
& \multicolumn{5}{|c}{b=1.3 \, fm^{-1}, \: N=4 }\\
 & & & & & & \\
\hline
 & & & & & & \\
j & a_j & i & Z_{N, \, i} & \lambda_i & j & a_j\\
 & (fm^{-1}) & & & (fm^{-2}) & & (fm^{-1})\\
\hline
  \begin{array}{c}
\\ 1\\ 2\\ \\ \\ \\ \\ \\  \\
  \end{array}
 & \begin{array}{l}
\\2.276012669\\ .040033331\\ \\ \\ \\ \\ \\ \\
   \end{array}
 & \begin{array}{c}
\\ 1\\ 2\\ 3\\ 4\\ \\ \\ \\  \\
  \end{array}
 & \begin{array}{l}
\\.1493428930\\ .4054072736\\ .6619688746\\ .6124857973\\ \\ \\ \\ \\
   \end{array}
 &\begin{array}{l}
\\.07258480091\\ .6661380111\\ 3.203427534\\ 37.\\ \\ \\ \\ \\
   \end{array}
 & \begin{array}{c}
\\1\\ 2\\ 3\\ 4\\ 5\\ 6\\ 7\\ 8\\
  \end{array}
 & \begin{array}{l}
\\3.552401289+7.346450796\,\mbox{i}\\ 3.552401289-7.346450796\,\mbox{i}\\
.8278500631+1.088427930\,\mbox{i}\\ .8278500631-1.088427930\,\mbox{i}\\
.5554972974+.5089227378\,\mbox{i}\\ .5554972974-.5089227378\,\mbox{i}\\
.4847238917\\ .04377880951\\
  \end{array}\\
\hline \hline
\end{array}
$$
}

\newpage
\subsection*{\begin{center} Figure captions \end{center}}

\subsubsection*{Figure 1}
The Bargmann $np$ potential for $^1S_0$ channel, phase equivalent to the
Yamaguchi potential \cite{Yamaguchi}.

\subsubsection*{Figure 2a}
The Bargmann $np$ potential for $^1S_0$ channel, phase equivalent to the
potential (\ref{PPot}) obtained by inverse scattering method in the $J$-matrix
approach \cite{ZInv}.

\subsubsection*{Figure 2b}
Same as Fig.~2a. The structure of second deep well of the $np$ Bargmann
potential.

\subsubsection*{Figure 2c}
Shallow structure of the $np$ Bargmann potential shown in Fig.~2a-b.

\subsubsection*{Figure 2d}
Same as Fig.~2c and asymptotic behaviour of the $np$ Bargmann potential.

\newpage
\includegraphics[bbllx=115,bblly=450,scale=1.3]{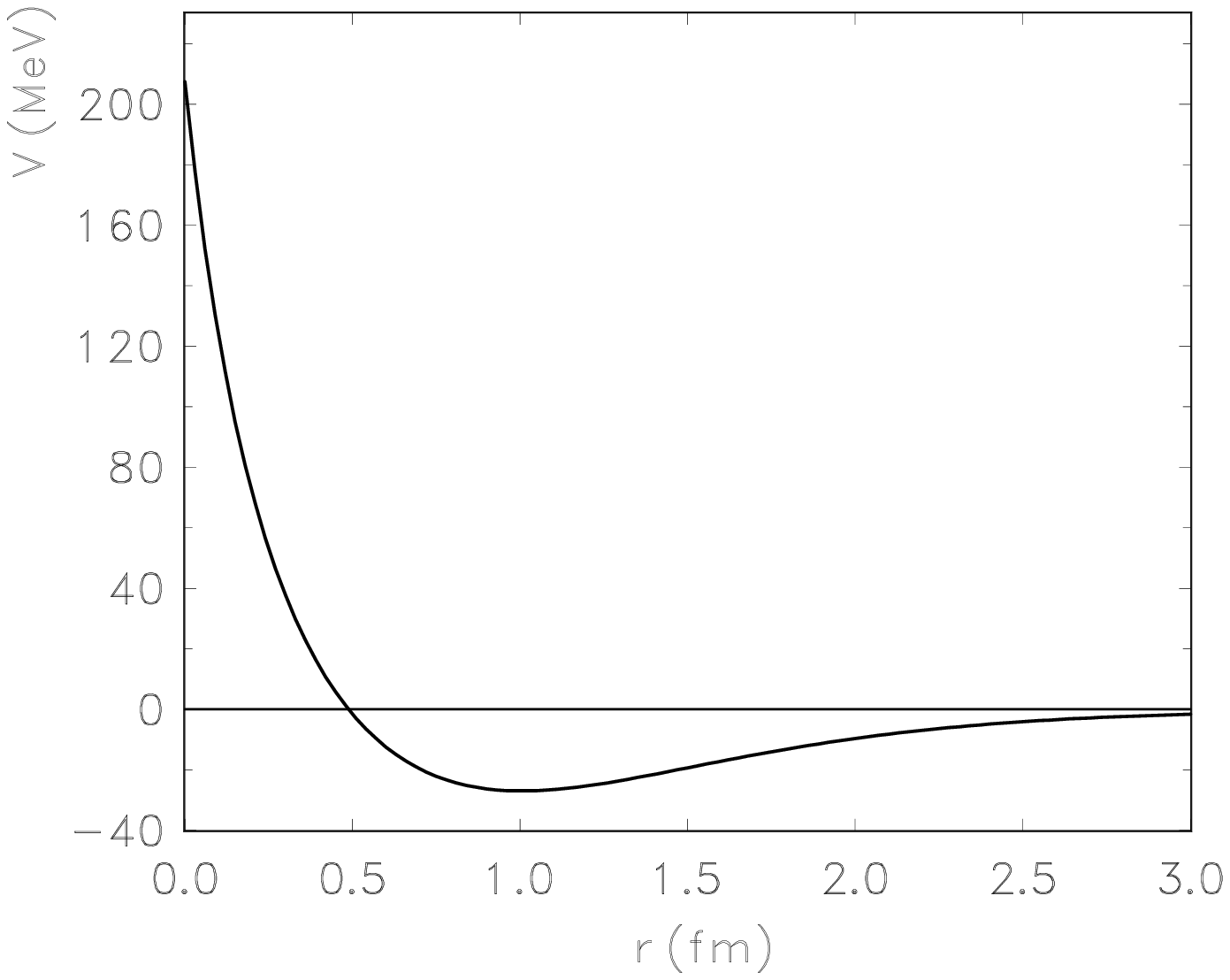}
\vfill
\begin{center}
{\Large \bf Figure 1.}
\end{center}

\newpage
\includegraphics[bbllx=110,bblly=450,scale=1.25]{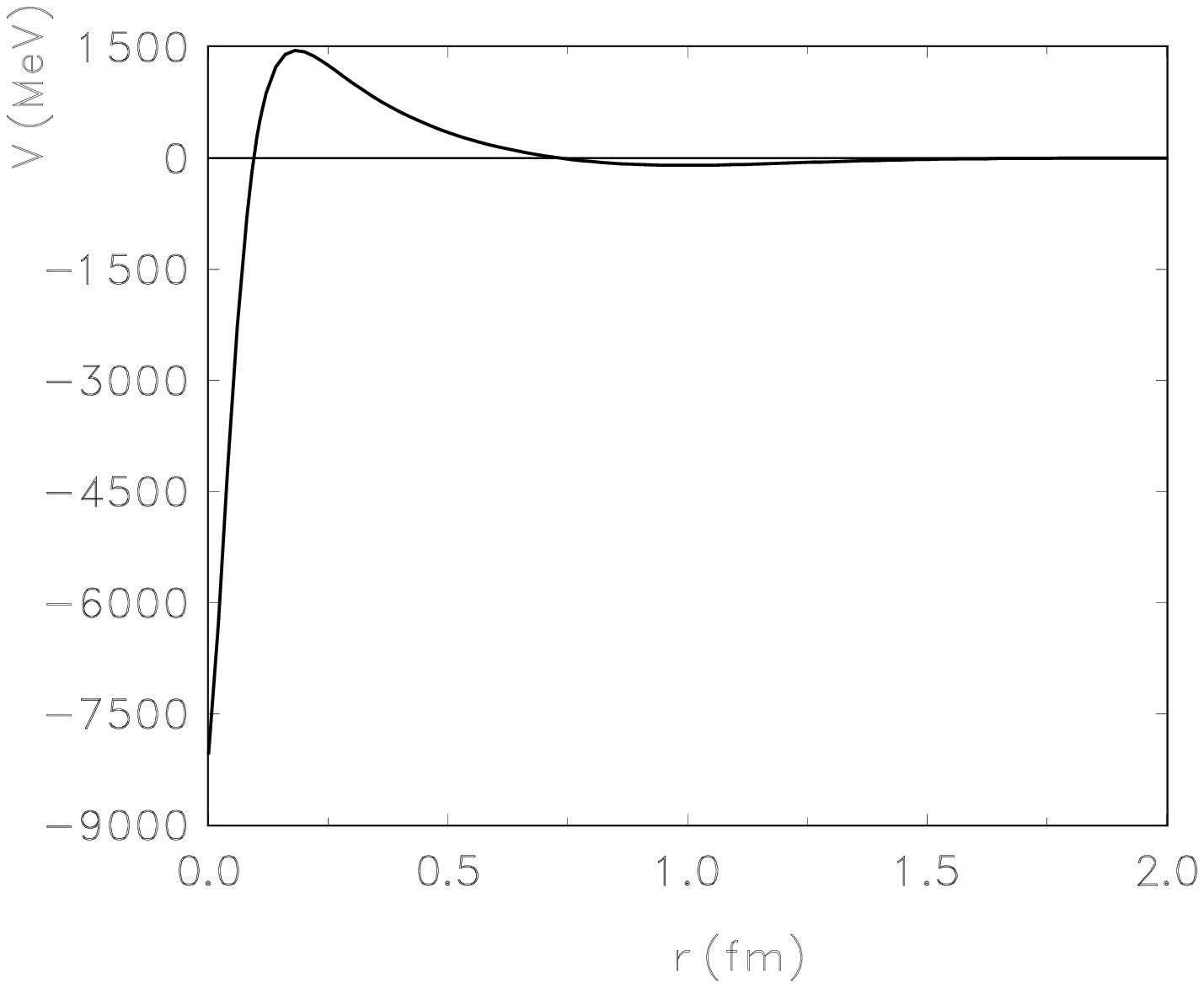}
\vfill
\begin{center}
{\Large \bf Figure 2a.}
\end{center}

\newpage
\includegraphics[bbllx=110,bblly=450,scale=1.25]{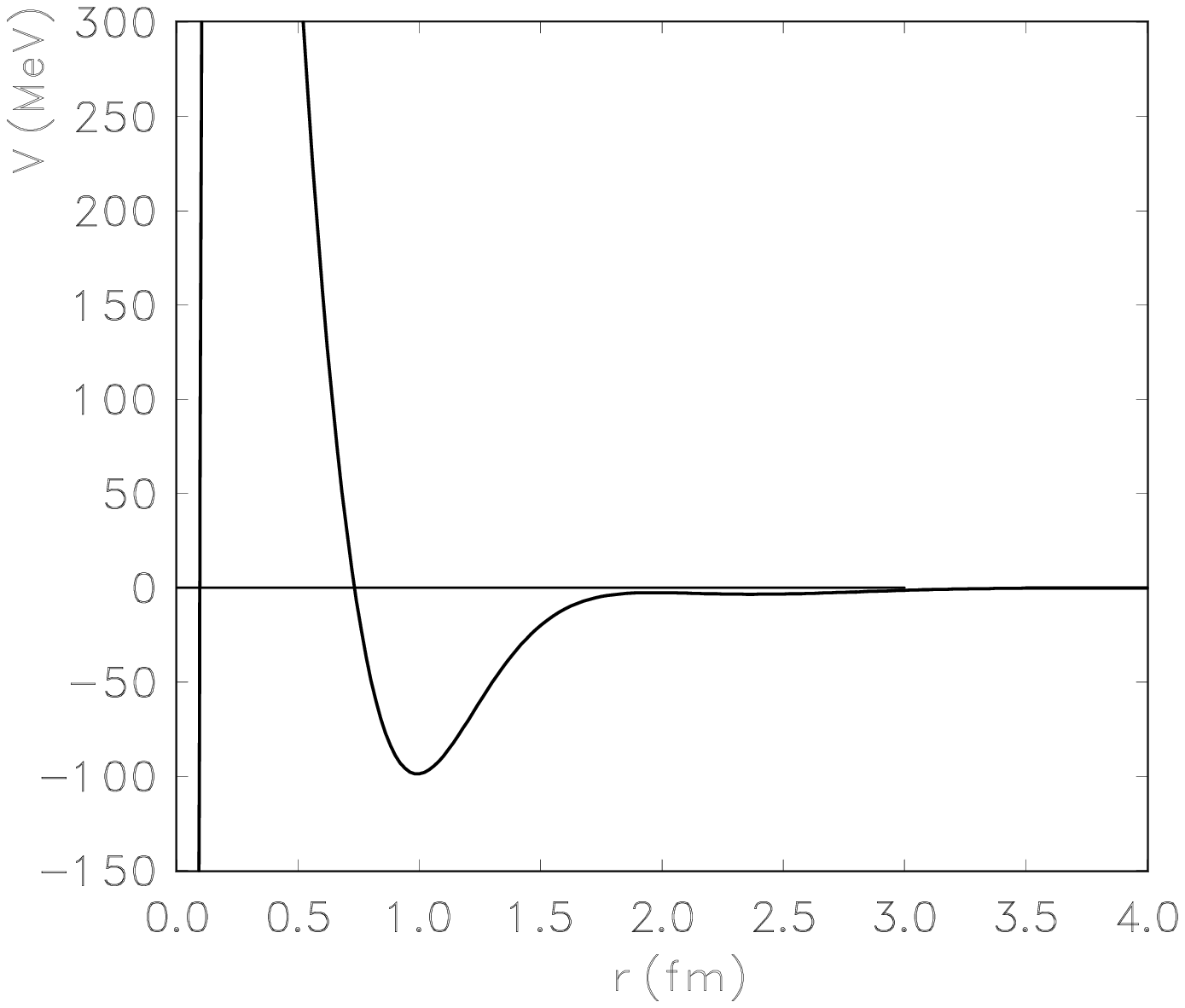}
\vfill
\begin{center}
{\Large \bf Figure 2b.}
\end{center}

\newpage
\includegraphics[bbllx=110,bblly=450,scale=1.25]{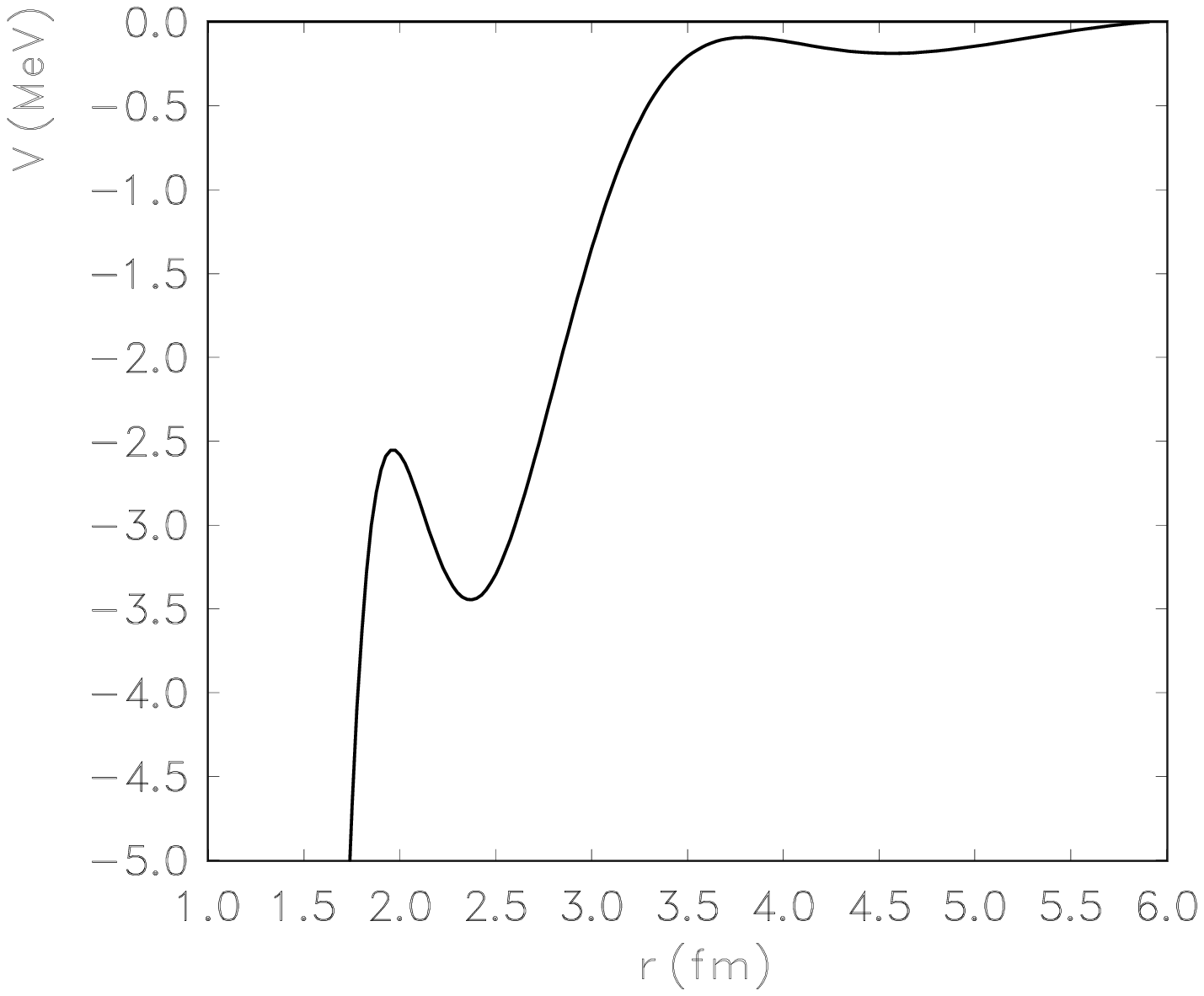}
\vfill
\begin{center}
{\Large \bf Figure 2c.}
\end{center}

\newpage
\includegraphics[bbllx=110,bblly=450,scale=1.25]{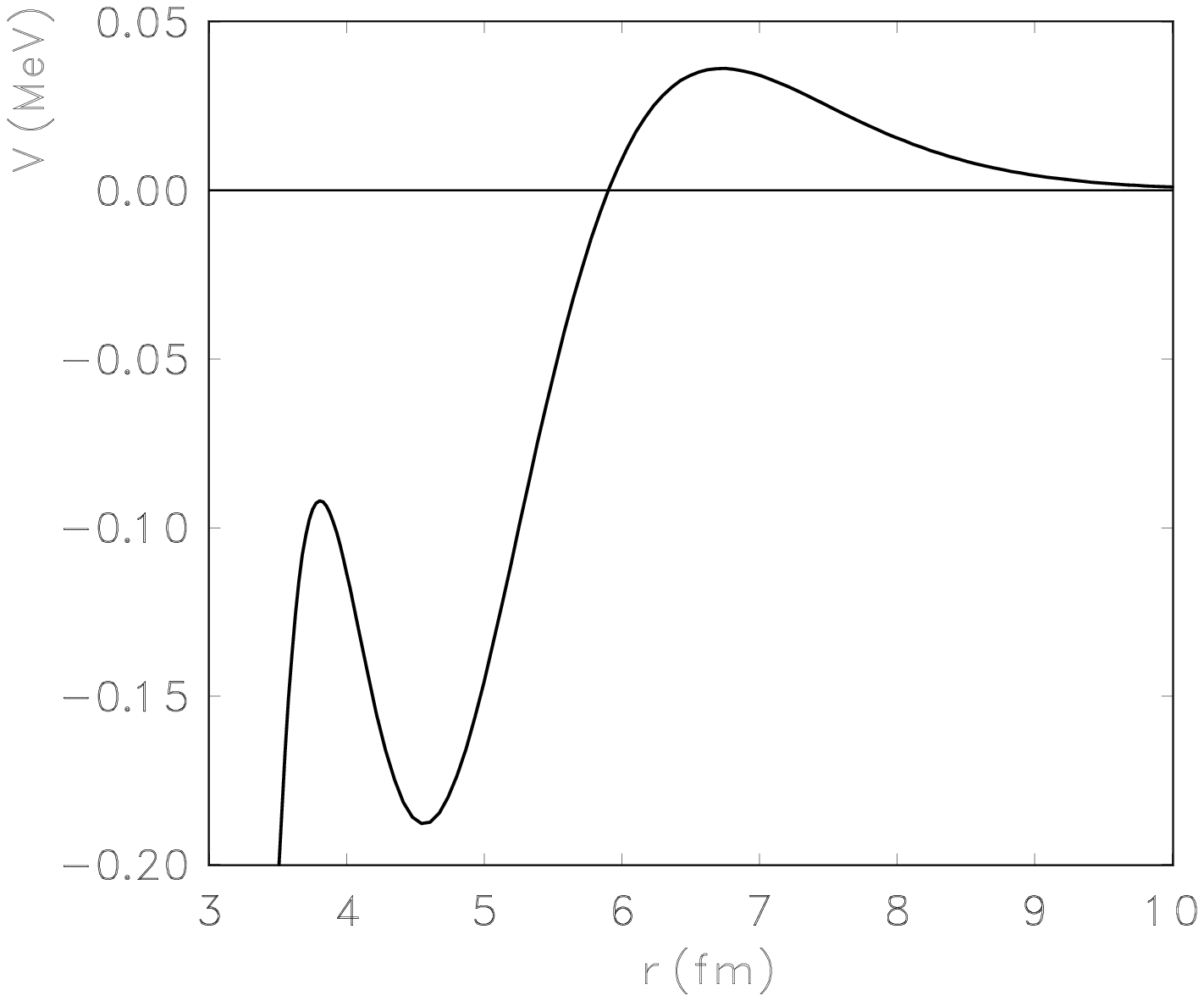}
\vfill
\begin{center}
{\Large \bf Figure 2d.}
\end{center}

\end{document}